\documentclass[namedreferences]{kluwer}  
\usepackage{graphicx}
\usepackage{amssymb}

\def\bbbr{\mathbb{R}}
\def\bbbt{\mathbb{T}}
\newcommand{\be}{\begin{equation}}
\newcommand{\ee}{\end{equation}}
\newcommand{\bea}{\begin{eqnarray}}
\newcommand{\eea}{\end{eqnarray}}
\newcommand{\crn}{\nonumber\\}

\newcommand{\centre}[2]{\multicolumn{#1}{c}{#2}}

\def\m@th{\mathsurround=0pt}
\def\EQM#1{\vcenter{\normalbaselines\m@th
    \ialign{${\displaystyle ##}$\hfil&&\ ${\displaystyle ##}$\hfil\crcr
    \mathstrut\crcr\noalign{\kern-\baselineskip}
    \noalign{\smallskip}
    #1\crcr\mathstrut\crcr\noalign{\kern-\baselineskip}}}}

\newcommand{\cS}{{\cal S}}
\newcommand{\SA}{\mathcal{S}\mathcal{A}\mathcal{B}\mathcal{A}}
\newcommand{\SB}{\mathcal{S}\mathcal{B}\mathcal{A}\mathcal{B}}
\newcommand{\C}{\gamma}
\newcommand{\g}{\gamma}
\renewcommand{\d}{\delta}
\newcommand{\dd}{\delta}
\newcommand{\cL}{{\cal L}}

\newcommand{\fl}{{ }}

\newcommand{\llabel}[1]{\hskip 1cm{\bf [\ #1\ ]}\label{#1}}
\renewcommand{\llabel}[1]{\label{#1}}  

\def\crm{\cr\noalign{\medskip}}

\def\trait{\noalign{\smallskip\hrule\smallskip}}

\def\Frac#1#2{{{\displaystyle\strut#1}\over{\displaystyle\strut#2}}}
\def\Dron#1#2{\Frac{\partial#1}{\partial#2}}

\def\Dt#1{\Frac{d#1}{dt}}
\def\etal{{\it et al.}}
\def\e{{\rm e}}
\newcommand{\ve}{\varepsilon} 
\renewcommand{\t}{\tau}

\newtheorem{Lem}{Lemma}
\newtheorem{prop}{Proposition}

\begin{document}
\begin{article}

\begin{opening}
\title{High order symplectic integrators \\ for
perturbed Hamiltonian systems}
\author{ Jacques Laskar and  Philippe Robutel}
\institute{Astronomie et Syst\`emes Dynamiques,\\ IMC-CNRS EP1825,\\
77 Av Denfert-Rochereau, 75014 Paris} 
\date{April 27, 2000}

\runningtitle{High order symplectic integrators}
\runningauthor{J. Laskar and P. Robutel}

\begin{abstract}
We present a  class of  symplectic integrators  adapted for the integration
of perturbed Hamiltonian systems of the form $H=A+\ve B$. We 
give a  constructive proof that for 
all integer $p$, there exists an integrator with positive steps  
with a remainder of order $O(\tau^p\ve +\tau^2\ve^2)$, where $\tau$
is the stepsize of the integrator.
The analytical expressions  of  the leading terms of the remainders are given
at all orders. In many cases, a corrector step
can be performed such that the remainder becomes $O(\tau^p\ve +\tau^4\ve^2)$.
The performances of these integrators are compared for the simple pendulum and 
the planetary $3$-Body problem of Sun-Jupiter-Saturn.
\end{abstract}

\end{opening} 

\section{Introduction} 
Symplectic integrators, due to their good stability properties are now currently
used for long time integrations of the Solar System, starting with the work of Wisdom and Holman (1991).
Despite some improvement resulting from  a good choice of initial
conditions (Saha and Tremaine, 1992) or a corrector to the output of the numerical integration (Wisdom
\etal, 1996), it is surprising that the integration method which is currently used 
in most computations (see  Duncan {\it et al.}, 1998) is  the celebrated 'leapfrog'  method
of order 2 (Ruth, 1983).
A reason for this choice is probably due to the fact that the methods of higher order 
which have been found by Forest and Ruth (1990) or Yoshida (1990) do not present very good 
stability properties for large  stepsize, due to the presence of negative 
steps.

In the present work, by considering perturbed Hamiltonians on the form $H=A + \ve B$ were both 
$A$ and $B$ are integrable,  we 
proove the existence  of a class of symplectic integrators with positive steps
 which   improve the precision of the integration by several order of magnitude with
respect to the  commonly used leapfrog method, and which present good stability properties at large
stepsize. 

\section{Lie formalism}

According to Yoshida, (1990), the search of symplectic integrators using Lie formalism was introduced
by Neri (1988). Since, it was largely developped by Yoshida (1990), Suzuki (1991, 1992), 
Koseleff (1993, 1996), and Mclachlan (1995, 1998).
Let  $H(p,q)$ be an Hamiltonian defined on  $\bbbr^n\times \bbbt^n$,
where  $(p,q)$ are the actions and angle variables. Hamilton equations are 
\be
\Dt{p_j} = -\Dron{H}{q_j} \ ; \quad  \Dt{q_j} = \Dron{H}{p_j}  
\ee
and the Poisson bracket of  $f, g$ is defined on  $\bbbr^n\times \bbbt^n$ by
\be
\{f,g\} = \sum_j \Dron{f}{p_j}\Dron{g}{q_j}- \Dron{f}{q_j}\Dron{g}{p_j}
\ee
If we denote  $x = (p,q)$, we obtain 
\be
\Dt{x} =  \{H,x\} = L_H x\ .
\label{eq.3}
\ee
where $L_H$ is the differential operator defined by
$
L_\chi\ f =  \{\chi,f\} 
$.
The solution $x(t)$ of (\ref{eq.3}) with $x(0)=x_0$ is obtained formally as
\be
x(t)= \sum_{n\geq 0}   \Frac{t^n}{n!}L_H^n \, x_0 =
 \e^{tL_H}\, x_0 \ . 
\ee

A symplectic scheme for integrating (\ref{eq.3}) from $t$ to $t+\tau$ consists  to approximate in a
symplectic way the operator 
$ \e^{\tau L_H}
$. Indeed, as $H=A+\ve B$, the Campbell-Baker-Hausdorf (CBH) theorem  ensures that 
\be
\e^{\t L_H} = \e^{\t L_A} \e^{\t L_{\ve B}} + o(\t) \ .
\ee
The operator 
$
S_1 = \e^{\t L_A} \e^{\t L_{\ve B}}
$
thus provides the most simple symplectic sheme for such Hamiltonians. 
This can be generalized with a combination of several steps involving successively $A$ and $\ve B$
in order to obtain integrators of higher orders. A general integrator with $n$ steps will be
\be
S_n =  \e^{c_1\t L_A} \e^{d_1\t L_{\ve B}}\cdots  \e^{c_n\t L_A} \e^{d_n\t L_{\ve B}}
\ee 
where the constants $(c_i,d_i)$ will be chosen in order to improve the order of the integrator. 
Using CBH theorem, and the linearity of the Lie derivative, we are ensured of the
existence of a formal series 
\be
\EQM{
\fl K &= k_{1,1} A + \ve k_{1,2}B  +\t\ve  k_{2,1} \{A,B\} \crm
\fl   &+\t^2\ve k_{3,1} \{A,\{A,B\}\} +\t^2\ve^2 k_{3,2}\{\{A,B\},B\}  \crm
\fl &+ \t^3\ve k_{4,1} \{A,\{A,\{A,B\}\}\} 
+ \t^3\ve^2 k_{4,2} \{A,\{\{A,B\},B\}\} \crm
&+ \t^3\ve^3 k_{4,3} \{\{\{A,B\},B\},B\} 
\fl + O(\t^4)
\label{eq.12}
}
\ee
where the coefficients $k_{i,j}$ are polynomials in the $(c_m,d_n)$, with rational coefficients, such
that
\be
S_n(\t) =  \e^{c_1\t L_A} \e^{d_1\t L_{\ve B}}\cdots  \e^{c_n\t L_A} \e^{d_n\t L_{\ve B}}= \e^{\t L_K}
\ee
It should be noted that in order to define these expressions in a non ambigous way, one needs to
decompose the Poisson brackets involving $A$ and $B$ over a basis of canonical elements of the
free  Lie
algebra $\cL(A,B)$ generated by $A$ and $B$ and the Poisson backet $\{\, ,\, \}$. Following Koseleff (1993), this is done here by using the Lyndon basis. The scheme
$S_n(\tau)$  integrates in an exact manner the formal Hamiltonian
$K$.  A symplectic integrator for $H = A+ \ve B$ will be obtained at order $p$  if 
$
K = A + \ve B + O(\t^p)
$. 
In the most general way, this will be achieved by solving the algebraic equations 
\bea
k_{1,1}=1 \ ; \qquad k_{1,2}=1 \ ; \crn
k_{i,j}=0 \qquad \hbox{for } \qquad (i\leq p) \, . 
\llabel{equa}
\eea 
In particular, we have 
$
k_{1,1} = c_1 + c_2 +\cdots c_n = 1 $, $
k_{1,2} = d_1 + d_2 +\cdots d_n = 1 $, for $p \geq 1$. 

\section{Symmetric integrators}
We will now restrict ourselves to symmetric integrators, that is 
integrators $S_n(\t)$ such that 
$
S_n(\t)^{-1} = S_n(-\t)
$.
We will  have 
\be
-\t L_{K(\t)} = -\t L_{K(-\t)}
\ee
thus $K(-\t) = K(\t)$, and the formal Hamiltonian $K(\t)$ is even.
As we distinguish $A$ and $\ve B$, we will have
several classes $\SA_k$ and $\SB_k$ of symmetric symplectic operators 
defined by their prototypes 
\be\EQM{
\fl\SA_{2n} : \e^{c_1\t L_A}\e^{d_1\t L_{\ve B}} \cdots 
    \e^{d_n\t L_{\ve B}}\e^{c_{n+1}\t L_A}\e^{d_n\t L_{\ve B}}
   \cdots \e^{d_1\t L_{\ve B}}  \e^{c_1\t L_A} \crm
\fl\SA_{2n+1} :\e^{c_1\t L_A}\e^{d_1\t L_{\ve B}}  \cdots 
     \e^{c_{n+1}\t L_A} \e^{d_{n+1}\t L_{\ve B}}\e^{c_{n+1}\t L_A}
   \cdots \e^{d_1\t L_{\ve B}}  \e^{c_1\t L_A} \crm
\fl\SB_{2n} : \e^{d_1\t L_{\ve B}}\e^{c_2\t L_A} \e^{d_2\t L_{\ve B}}\cdots 
    \e^{d_n\t L_{\ve B}}\e^{c_{n+1}\t L_A}\e^{d_{n}\t L_{\ve B}} 
   \cdots  \e^{c_2\t L_A} \e^{d_1\t L_{\ve B}} \crm
\fl\SB_{2n+1} : \e^{d_1\t L_{\ve B}}\e^{c_2\t L_A} \cdots 
       \e^{c_{n+1}\t L_A}\e^{d_{n+1}\t L_{\ve B}} \e^{c_{n+1}\t L_A}
   \cdots \e^{c_2\t L_A}\e^{d_1\t L_{\ve B}} \crm
}\ee
The index of the integrator is the number of evaluations of $A$ and $B$ which are necessary for
each step. With these notations, the classical leapfrog integrator can be 
considered as  $SBAB_1=\e^{\frac{\t}{2} L_{\ve B}}  \e^{\t L_A}  \e^{\frac{\t}{2} L_{\ve B}}\in\SB_1$
or as $SABA_1= \e^{\frac{\t}{2} L_A}\e^{\t L_{\ve B}} \e^{\frac{\t}{2} L_A}\in\SA_1$. 
 In both cases, the integrator is of order 2 and the formal Hamiltonian is 
$
K = A + \ve B + O(\t^2 \ve)
$.
The fourth order solution found by Forest and Ruth (1990) or in an other way by Yoshida (1990) is 
either of the form $\SA_3$ or $\SB_3$ that is, for $\SB_3$
\be
SFRA_3 = \e^{d_1\t L_{\ve B}} 
      \e^{c_2\t L_A} \e^{d_2\t L_{\ve B}}\e^{c_3\t L_A}\e^{d_2\t L_{\ve B}} \e^{c_2\t L_A}\e^{d_1\t L_{\ve B}} 
\ee
with
\be
\left\{\EQM{
c_3+2c_2 = 1 \cr
d_1+d_2=1/2 \cr
1/12 -1/2\,c_2 +1/2\,c_2^2 +c_2\,d_1 -c_2^2\,d_1 =0 ;\cr
-1/24 +1/4\,c_2 -c_2\,d_1 +c_2\,d_1^2 =0
}\right.
\ee
This system has a single real solution with  approximate values 
$d_1 \approx   0.6756$,
$c_2  \approx   1.3512$,
$d_2  \approx -0.1756$,
$c_3  \approx -1.7024$.
The problem with this integrator, is that due to the presence of negative time steps, the absolute
value of the time steps remains high, and for large stepsizes, at an equivalent cost, the 
leapfrog integrators becomes more effective.
In fact, Suzuki (1991) has demonstrated that it is not possible to obtain integrators
of order $p> 2$ with only
positive steps. The problem of the negative stepsize 
can nevertheless be overcome.
 
\section{Integrators for perturbed Hamiltonian}
In the previous sections, we have not yet taken into account the existence of 
the small parameter $\ve$.   Indeed, the terms of second order of $K$ (\ref{eq.12}) are 
$\t^2\ve k_{3,1} \{A,\{A,B\}\}$ and  $ \t^2\ve^2 k_{3,2}\{\{A,B\},B\} $ which are respectively of order $\t^2\ve$
and $\t^2\ve^2$. One can thus try to cancel uniquely the largest term, that is 
$
 k_{3,1} = 0  
$.
This can be done using 
\be
\SA_{2} : \e^{c_1\t L_A} \e^{d_1\t  L_{\ve B}} \e^{c_2\t L_A}  \e^{d_1\t  L_{\ve B}} \e^{c_1\t L_A}
\label{eq.SA2} 
\ee
or 
\be
\SB_{2} : \e^{d_1\t  L_{\ve B}}\e^{c_2\t L_A}  \e^{d_2\t  L_{\ve B}} \e^{c_1\t L_A}  \e^{d_1\t  L_{\ve B}}  .
\label{eq.SB2}
\ee

With the type  $\SA_2$, one obtains
$
d_1 = \frac{1}{2}$, $c_2 = 1-2c_1 $
and 
\be
\EQM{
K_{\SA_2} =  A + \ve B &+\t^2\ve(\Frac{1}{12} -\Frac{1}{2}c_1 +\Frac{1}{2}c_1^2) \{A,\{A,B\}\} \crm
&+
\t^2\ve^2( -\Frac{1}{24}+\Frac{1}{4}c_1) \{\{A,B\},B\} + O(\t^4\ve)
}
\ee
As we search for only positive stepsize, we find a unique solution for cancelling the term 
in $\ve\t^2$, that is
\be
 c_2 = \Frac{1}{\sqrt{3}}\, ;\quad c_1 = \Frac{1}{2}(1-\Frac{1}{\sqrt{3}})\, ;\quad d_1 = \Frac{1}{2} \,
;
\ee
with these coefficients, we obtain 
$
K_{\SA_2} =  A + \ve B +O(\t^4\ve + \t^2\ve^2)
$.
In a similar way, we obtain the solution for $\SB_2$
\be
 d_2 = \Frac{2}{3}\, ;\quad d_1 = \Frac{1}{6}\, ;\quad c_2 = \Frac{1}{2} \,
;
\ee
and as previously 
$
K_{\SB_2} =  A + \ve B +O(\t^4\ve + \t^2\ve^2)
$.
Quite surprisingly, this latest integrator which  is in most cases much more precise than the leapfrog integrator
($\SB_1$)  at the same cost (see section 8), does not seem to have been used so far. 

\section{Higher orders}
It becomes then tempting to iterate this process at higher order. We will not try to remove the term 
of order $\t^2\ve^2$, which is not the most important for large stepsize when $\ve$ is small. We will
search for solutions $S_n$ of the form $\SA_n$ or $\SB_n$ for which the associated Hamiltonian $K_{S_n}$
verifies  
\be
K_{S_n} = A + \ve B +O(\t^{2n}\ve + \t^2\ve^2)
\ee
For this, we need to cancell
 at all order $p < 2n $ the coefficient  $k_{p,1}$ of the single term
of order $\t^p\ve$ in the Lyndon decomposition of $K_{S_n}$
\be
\t^p\ve k_{p,1}\{A,\{A,\{A,\dots \{A,B\}\}\}\dots\}
\ee
We thus need to compute the part of  $K_{S_n}$ which is of degree $\leq 1$  in $B$. 
We will use some 
results on calculus on free Lie algebra for which the reader 
should refer to (Bourbaki, 1972). We will call $\cL(U,V)$
 the free Lie algebra generated by $U$ and $V$, endowed with its canonical 
 associative structure.
We will also use  the symbol $\equiv$ for 
the equality in $\cL(U,V)$ modulo terms of degree $\geq 2$ in $V$. We have the two lemmas (Bourbaki, 1972)
\begin{Lem}
\be
\e^{U}\,V\,\e^{-U} = \e^{ad(U)}\, V
\ee
\label{lem1}
\end{Lem}
where the exponential of  $X$ is formally defined as
$
\exp(X) = \sum_{n=0}^{+\infty} {X^n}/{n!} 
$, 
and where the adjoint operator $ad$ is defined as $ad(X).Y = [X,Y] $.
\begin{Lem}
\be
\e^{U+V} \equiv \e^U + \e^U\left(\Frac{1-\e^{-ad(U)}}{ad(U)} \right)\, V  \ .
\ee
\label{lem2}
\end{Lem}
The next result is a generalisation of a
classical  expansion at degree 1 in V of the Campbell-Baker-Haussdorff formula.

\begin{prop}
Let $\C \in \bbbr$. Then there exists $W \in \cL(U,V)$ such that
\be
\e^{\C U}\e^V \e^{(1-\C) U}= \e^W
\ee
\noindent with
\be
W \equiv U +  \Frac{ad(U)\e^{\C \,ad(U)}} {\e^{ad(U)}-1} \, V
\ee
\noindent that is
\be
W \equiv U  + \sum_{p=0}^{+\infty} \Frac{B_p(\C)}{p!} ad(U)^p\, V  
\ee
and
where  $B_n(x)$ are the Bernoulli polynomial defined as
\be
\Frac{t\,\e^{tx}}{\e^t-1} = \sum_{n=0}^{+\infty} B_n(x) \Frac{t^n}{n!}
\ee\end{prop}
Indeed, the existence of $W \in \cL(U,V) $ satisfying the above relation is 
given by the CBH  theorem, on the other hand,  we have
\be
\e^{ \C U}\e^V \e^{(1-\C) U} \equiv \e^U + \e^U\,\e^{(\C-1) U}V \e^{(1-\C) U}
\ee
and from lemma \ref{lem1}, this is also equal to 
\be
\e^U + \e^U\,\e^{(\C-1) ad(U)}V \ .
\ee
As for $V=0$, we have $W=U$, we can set $W  \equiv U + W_1 $, 
where $W_1$ is  of degree 
1 in $V$, and from lemma \ref{lem2}
\be
\e^W \equiv \e^U + \e^U\left(\Frac{1-\e^{-ad(U)}}{ad(U)} \right)\, W_1
\ee
thus
\be
W_1 = \Frac{ad(U)\e^{(\C-1) ad(U)}} {1-\e^{-ad(U)}} \, V
\ee
which ends the proof. For $\C=1$, we recover the CBH results.
This result is  then easily generalized to the case of multiple products.

\begin{prop}
Let $c_1,\dots c_n, d_1,\dots, d_n \in \bbbr$, such that $\sum_{i=1}^n c_i = 1$. Then there exists $W \in \cL(U,V)$ such that
\be
\e^{c_1 U}\e^{d_1 V} \e^{c_2 U}\e^{d_2 V} \dots \e^{c_n U}\e^{d_n V}= \e^W
\label{eq.gen_form}
\ee
\noindent with
\be
W \equiv U + \sum_{k=1}^n d_k  \Frac{ad(U)\e^{\C_k \,ad(U)}} {\e^{ad(U)}-1} \, V
\ee
\noindent that is
\be 
W \equiv U  +\sum_{p=0}^{+\infty} \left( \sum_{k=1}^n d_k \,\Frac{B_p(\C_k)}{p!}\right) ad(U)^p\, V  
\ee
with $\C_k = c_1+\dots +c_k$.
\end{prop}
Dems. This is staightforward as soon as we remark that 
\be
\e^{c_1 U}\e^{d_1 V} \e^{c_2 U}\e^{d_2 V} \dots \e^{c_n U}\e^{d_n V} \equiv \sum_{k=1}^n d_k 
\e^{\C_k U}\,V \e^{(1-\C_k) U} + \e^U
\ee
Remark : As $B_0(x)=1$, if $\sum_{k=1}^n d_k=1$, we have 
\be
W \equiv U  + V + \sum_{p=1}^{+\infty} \left(\sum_{k=1}^n d_k \,\Frac{B_p(\C_k)}{p!} \right) ad(U)^p\, V   \ .
\ee
\section{Computation of the coeffcients}
Propostion 2  , applied  with $U= \tau L_A $ and $V=\tau \ve L_B$, gives 
directly  the algebraic equations which could then be 
solved for obtaining integrators of  arbitrary order for perturbed systems. The problem 
is thus reduced to the search for coeffcients $\C_k, d_k$ such that 
\be
 \sum_{k=1}^n d_k\,g(\C_k,t) = 1 + o(t^N)
 \label{eq.cond}
\ee
for $N$ as high as possible with
\be
g(x,t) =  \Frac{t\, \e^{x\,t}}{\e^t-1}  \ .
\ee
That is, with $\sum_{k=1}^n c_k=1 $, we will have to solve an algebraic system of equations  of the form
\be
\EQM{
\sum_{k=1}^n d_k\,B_0(\C_k) = \sum_{k=1}^n d_k = 1 \cr
\sum_{k=1}^n d_k\,B_p(\C_k) = 0 \quad \hbox{for $0 < p \leq N $} \ .
\label{eq.alg}
}
\ee
It should be noted that all the integrators $\SA_n$ and  $\SB_n$ can be written on the general form
(\ref{eq.gen_form}) by taking $d_n= 0$ or $c_1=0$ in (\ref{eq.gen_form}).
Moreover, if we search for symmetric integrators, all the relations 
in (\ref{eq.alg}) will be automatically fullfilled for  odd values of $p$. In this case, we just have to consider
even values of $p$, for which we give  the Bernoulli polynomials up to $p=10$.
\be
\EQM{B_0(x) = 1 \cr
B_2(x) = \frac{1}{6} - x + x^2  \cr
B_4(x) = -\frac{1}{30}  + x^2 - 2\, x^3 + x^4  \cr
B_6(x) = \frac{1}{42} - \frac{x^2}{2} + \frac{5\, x^4}{2} - 3\, x^5 + x^6  \cr
B_8(x) = -\frac{1}{30}  + \frac{2\, x^2}{3} - \frac{7\, 
      x^4}{3} + \frac{14\, x^6}{3} - 4\, x^7 + x^8  \cr
B_{10}(x) = \frac{5}{66} - \frac{3\, x^2}{2} + 5\, x^4 - 7\, 
x^6 + \frac{15\, x^8}{2} - 5\, x^9 + x^{10}
}\ee
 
 For example, the first integrators  
 $SABA_2 =  \e^{c_1 U}\e^{d_1 V}  \e^{c_2 U} \e^{d_1 V}  \e^{c_1 U}$
  will be obtained by solving the set of equations
 \be
 \EQM{
 2c_1+c_2 &= 1\cr
 2 d_1 &= 1  \cr
 d_1 B_2(\C_1) + d_1 B_2(\C_2) &= 0
 }
 \ee
 with $\C_1 =c_1, \C_2 = c_1+c_2 $, thus  $\C_2 = 1-\C_1$. As $ g(1-x,t)= g(x,-t)$, we have for all $p$ 
 \be
 B_p(1-x) = (-1)^p B_p(x)
 \ee 
 and the previous system reduces to
 \be
 d_1 = 1/2 \qquad c_2 = 1-2c_1 \qquad B_2(c_1) = 0
 \ee
 and we recover the previous results.  For $SABA_3 =  \e^{c_1 U}\e^{d_1 V}  \e^{c_2 U}  \e^{d_2 V} \e^{c_2 U}\e^{d_1 V}  \e^{c_1 U}$
we have
 \be
 \EQM{
 c_1+c_2 &= 1/2\cr
 d_2 +2 d_1 &= 1  \cr
 d_1 B_2(\C_1) + d_2 B_2(\C_2) + d_1 B_2(\C_3) &= 0 \cr
 d_1 B_4(\C_1) + d_2 B_4(\C_2) + d_1 B_4(\C_3) &= 0 \cr
}
 \ee
with $\C_1 = c_1$, $\C_2=c_1+c_2=1/2$, and $\C_3 = c_1+c_2+c_2 = 1-c_1$. We have thus 
$B_2(\C_2) = -1/12, B_4(\C_2)=7/240, B_2(\C_3)=B_2(c_1), B_4(\C_3)=B_4(c_1)$. We are thus left with
\be
\EQM{
c_2 = 1/2-c_1 \cr
d_2 = 1-2 d_1 \cr
 d_1 B_2(c_1) - (1-2d_1)/24 = 0 \cr
 d_1 B_4(c_1) + 7(1-2d_1)/480 = 0 \cr
}
\ee
The resolution of this system is made easily and provide a single solution for which all the 
coefficients $c_i, d_i$ are positive
\be
c_1 = {\frac{5 - {\sqrt{15}}}{10}} ;\quad c_2 =\frac{\sqrt{15}}{10} ; \quad d_1 = \frac{5}{18} ; \quad d_2 = \frac{4}{9}
\ee

This can be continued at all orders, but  algebraic equations becomes more complicated as the order increases. 
The symplectic integrators up to  order 10 are listed in Table I.

{\scriptsize\renewcommand\arraystretch{0.7}

\def\B#1{\e^{d_{#1}B}}
\def\A#1{\e^{c_{#1}A}}
 
\begin{table}[h]
\label{table_a}
\caption{Coefficients of the integrators 
$\SA_n$ and $\SB_n$  up to $n=10$.}
\begin{tabular}{cc|cc}
\trait 
\centre{4}{$SABA_1 $}\\ 
\trait
$c_1$&  ${1}/{2} $   & $d_1$  & $1$\\
\trait 
\centre{4}{$SABA_2  $}\\
\trait
$c_1$& $1/2-\sqrt{3}\,/6 $    & $d_1$ &$ {1}/{2} $ \\
$c_2$& $\sqrt{3}\, /3$    &  &\\
\trait 
\centre{4}{$SABA_3  $}\\
\trait
$c_1$& $1/2-\sqrt{15}\,/10 $ & $d_1$ &$ {5}/{18} $ \\
$c_2$& $\sqrt{15}\,/10 $    & $d_2$ &$ {4}/{9}$\\ 
\trait 
\centre{4}{$SABA_4 $}\\
\trait
$c_1$& $1/2 -\sqrt{525+70\sqrt{30}}\,/70 $   & $d_1$ &  $ 1/4 -  \sqrt{30}\,/72$\\
$c_2$& $ \left(\sqrt{525+70\sqrt{30}} -\sqrt{525-70\sqrt{30}}\right)/70 $    & $d_2$ &$1/4 +  \sqrt{30}\,/72$ \\
$c_3$& $ \sqrt{525-70\sqrt{30}}\, /35$    &  &\\
\trait 
\centre{4}{$SABA_5 $}\\
\trait
$c_1$&  $1/2- (\sqrt{490+42\sqrt{105}} +  \sqrt{490-42\sqrt{105}})/84$   & $d_1$ & $(322-13\sqrt{70})/1800$\\
$c_2$&  $\sqrt{490-42\sqrt{105}}\,/42 $ & $d_2$ & $(322+13\sqrt{70})/1800$ \\
$c_3$& $(\sqrt{490+42\sqrt{105}} -  \sqrt{490-42\sqrt{105}})/84$   & $d_3$ & 64/225\\
\trait 
\centre{4}{$SABA_6$}\\
\trait
$c_1$&  $0.033765242898423986093849222753002695$   &$d_1$&$0.085662246189585172520148071086366447$  \\
$c_2$&  $0.135630063868443757075450979737044631$   &$d_2$&$0.180380786524069303784916756918858056$\\
$c_3$&  $0.211295100191533802515448936669596706$   &$d_3$&$0.233956967286345523694935171994775497$\\
$c_4$&  $0.238619186083196908630501721680711935$   &  &\\
\trait 
\centre{4}{$SABA_7 $}\\
\trait
$c_1$&  $0.025446043828620737736905157976074369$   &$d_1$&$0.064742483084434846635305716339541009$  \\
$c_2$&  $0.103788363371682042331162455383531428$   &$d_2$&$0.139852695744638333950733885711889791$  \\
$c_3$&  $0.167843017110998636478629180601913472$   &$d_3$&$0.190915025252559472475184887744487567$  \\
$c_4$&  $0.202922575688698583453303206038480732$   &$d_4$&$256/1225 $  \\
\trait 
\centre{4}{$SABA_8 $}\\
\trait
$c_1$&  $0.019855071751231884158219565715263505$   &$d_1$&$0.050614268145188129576265677154981095$  \\
$c_2$&  $0.081811689541954746046003466046821277$   &$d_2$&$0.111190517226687235272177997213120442$  \\
$c_3$&  $0.135567033748648876886907443643292044$   &$d_3$&$0.156853322938943643668981100993300657$  \\
$c_4$&  $0.171048883710339590439131453414531184$   &$d_4$&$ 0.18134189168918099148257522463859781$  \\
$c_5$&  $0.183434642495649804939476142360183981$   &&  \\
\trait 
\centre{4}{$SABA_9$}\\
\trait
$c_1$&  $0.015919880246186955082211898548163565$   &$d_1$&$0.040637194180787205985946079055261825$  \\
$c_2$&  $0.066064566090495147768073207416968997$   &$d_2$&$0.090324080347428702029236015621456405$  \\
$c_3$&  $0.111329837313022698495363874364130346$   &$d_3$&$0.130305348201467731159371434709316425$  \\
$c_4$&  $0.144559004648390734135082012349068788$   &$d_4$&$0.156173538520001420034315203292221833$  \\
$c_5$&  $0.162126711701904464519269007321668304$   &$d_5$&$16384/99225 $  \\
\trait 
\centre{4}{$SABA_{10}$}\\
\trait
$c_1$&  $0.013046735741414139961017993957773973$   &$d_1$&$0.033335672154344068796784404946665896$  \\
$c_2$&  $0.054421580914093604672933661830479502$   &$d_2$&$0.074725674575290296572888169828848666$  \\
$c_3$&  $0.092826899194980052248884661654309736$   &$d_3$&$0.109543181257991021997767467114081596$  \\
$c_4$&  $0.123007087084888607717530710974544707$   &$d_4$&$0.134633359654998177545613460784734677$  \\
$c_5$&  $0.142260527573807989957219971018032089$   &$d_5$&$0.147762112357376435086946497325669165$  \\
$c_6$&  $0.148874338981631210884826001129719985$   && \\
\trait
\end{tabular}
\end{table}

\begin{table}[h]
\begin{tabular}{cc|cc} 

\trait 
\centre{4}{$SBAB_1 $}\\
\trait
$c_2$ & $1$    & $d_1$ &$1/2 $ \\ 
\trait 
\centre{4}{$SBAB_2  $}\\
\trait
& & $d_1$ &$1/6$ \\
$c_2$ & $1/2 $    & $d_2$ & $2/3$\\
\trait 
\centre{4}{$SBAB_3  $}\\
\trait
$c_2$ & $1/2-\sqrt{5} \,/10$                              & $d_1$ &  $1/12$\\
$c_3$ & $\sqrt{5}\,/5$      & $d_2$ &$5/12$ \\
\trait 
\centre{4}{$SBAB_4 $}\\
\trait
& & $d_1$ &  $1/20$\\ 
$c_2$ & $ 1/2-\sqrt{3/7}\,/2 $     & $d_2$ & $49/180$ \\
$c_3$ & $  \sqrt{3/7} \, /2 $  & $d_3$ & 16/45\\
\trait 
\centre{4}{$SBAB_5$}\\
\trait
 $c_2$ & $1/2-\sqrt{3+6/\sqrt{7}}\,/6$     & $d_1$ & $1/30$  \\
 $c_3$ & $ (\sqrt{3+6/\sqrt{7}}-\sqrt{3-6/\sqrt{7}} )\,/6$    & $d_2$ &  $ (14-\sqrt{7})/60 $\\
$c_4$ & $\sqrt{1/3-{2}/{3\sqrt{7}}}$   & $d_3$ &$(14+\sqrt{7})/60$ \\
\trait 
\centre{4}{$SBAB_6$}\\
\trait
 &    & $d_1$ & $1/42$  \\
$c_2$ & $1/2-\sqrt{(15+2 \sqrt{15})/33}\,/2$    & $d_2$ &  $ 31/175-\sqrt{3/5}\,/20 $\\
$c_3$ & $\sqrt{5/22 -\sqrt{5/33}/2}$    & $d_3$ &$ 31/175+\sqrt{3/5}\,/20 $ \\
$c_4$ &$\sqrt{5/44 -\sqrt{5/3}/22}$   & $d_4$ & $128/525$\\
\trait 
\centre{4}{$SBAB_7$}\\
\trait
$c_2$ &  $0.064129925745196692331277119389668281$   &$d_1$&$1/56$  \\
$c_3$ &  $0.140019983538232156596467514911355124$   &$d_2$&$0.105352113571753019691496032887878162$  \\
$c_4$ &  $0.191200481765331716687926735526300967$   &$d_3$&$0.170561346241752182382120338553874086$\\
$c_5$ &  $0.209299217902478868768657260345351255$   &$d_4$&$0.206229397329351940783526485701104895$\\
\trait 
\centre{4}{$SBAB_8 $}\\
\trait
$   $ &  $ $   &$d_1$&$1/72 $  \\
$c_2$ &  $0.050121002294269921343827377790831021$   &$d_2$&$0.082747680780402762523169860014604153$  \\
$c_3$ &  $0.111285857950361201933229908663497754$   &$d_3$&$0.137269356250080867640352809289686363$  \\
$c_4$ &  $0.157034407842279797367566679191341619$   &$d_4$&$0.173214255486523172557565766069859144$  \\
$c_5$ &  $0.181558731913089079355376034354329607$   &$d_5$&$2048/11025 $  \\
\trait 
\centre{4}{$SBAB_9 $}\\
\trait
$c_2$ &  $0.040233045916770593085533669588830933$   &$d_1$&$1/90 $  \\
$c_3$ &  $0.090380021530476869412913242981253705$   &$d_2$&$0.066652995425535055563113585377696449$  \\
$c_4$ &  $0.130424457647530289670965541064286364$   &$d_3$&$0.112444671031563226059728910865523921$  \\
$c_5$ &  $0.156322996072028735517477663386542232$   &$d_4$&$0.146021341839841878937791128687221946$  \\
$c_6$ &  $0.165278957666387024626219765958173533$   &$d_5$&$0.163769880591948728328255263958446572$  \\
\trait 
\centre{4}{$SBAB_{10} $}\\
\trait 
$ $ &  $ $   &$d_1$&$1/110 $  \\
$c_2$ &  $0.032999284795970432833862931950308183$   &$d_2$&$0.054806136633497432230701724790175355$  \\
$c_3$ &  $0.074758978372457357854928159995462766$   &$d_3$&$0.093584940890152602054070760949717460$  \\
$c_4$ &  $0.109624073333469706075726923315353220$   &$d_4$&$0.124024052132014157020042433210936377$  \\
$c_5$ &  $0.134738595704632807519526226959347078$   &$d_5$&$0.143439562389504044339611201665767616$  \\
$c_6$ &  $0.147879067793469695715955757779528754$   &$d_6$&$32768/218295 $  \\
\trait 
\end{tabular}
\end{table} 
}
\clearpage

\section{McLachlan  solution}
While we were writing a first version of this work, 
we realized that McLachlan (1995) had already found all 
the previous integrators.  The paper of McLachlan is obviously not well-known to astronomers,
otherwise they would have used at least the integrators $\SA_2,\SB_2,\SA_3$ and $\SB_3$ which 
have very good properties\footnote{The first integrators of the family ($\SA_2,\SB_2,\SA_3$ and $\SB_3$) 
have been 
also recently reported by Chambers and Murison (2000). 
The integrator $\SB_2$ is mentioned in the book of E. Forest (1998).}.  McLachlan  just makes the computations 
on a very simple example for which  the  integration of the equations reduces to a simple 
integral.
He then claims that this is representative of the most general case. 
Although  this  may be  true, the argument is not as straightforward as 
the constructive method which is presented here.
On the other hand, the final remarks of McLachlan (1995) can be adapted here to complete the present proof 
and to provide the expression for the coefficients of these symplectic  integrators at any order.
Indeed, if we observe that 
\be
\Frac{\e^t - 1}{t} = \int_0^1 \e^{xt} dx \ ,
\ee
and that $(\e^t - 1)/{t} = O(1) $, the problem of finding $d_k, \C_k$ verifying (\ref{eq.cond}) is equivalent 
to the search of weights $d_k$ and nodes $\C_k$ such that 
\be
\sum_{k=1}^n d_k\,\e^{\C_k\,t} = \int_0^1 \e^{xt} dx + o(t^N)
\label{eq.66}
\ee
The solution of this problem is  known classically as the Gauss integration formula. The 
values of $\C_k$ are given by $\C_k = (1+x_k)/2$ where $x_k$ 
are the roots of the  degree $n$ Legendre polynomial 
$P_n(x)$. The  the associated weights $d_k$ are all positive and are given by
\be
d_k = \Frac{1}{(1-x_k^2)\left(P'_n(x_k)\right)^2}
\ee

More precisely, if we consider an integrator of type
\be
\SA_n : \e^{c_1 U}\e^{d_1 V} \e^{c_2 U}\e^{d_2 V} \dots \e^{c_n U}\e^{d_n V}\e^{c_{n+1} U} \ ,
\ee without any assumption of 
 symmetry, we will have, in the above formula $d_{n+1} = 0$,  thus, for $ k = 1,\dots,n$, the coefficients 
  $ \C_k =  (1+x_k)/2$ where $x_k$  are the roots of $P_n(x)$. All $x_k$ are in the interval $[-1,1]$. We will thus have 
  $\C_k \in [0,1]$. If we put the $\C_k$ in ascending order, the values of the 
  coefficients $c_k = \C_{k+1}-\C_k$ are  all positive and $c_{n+1} = 1-\C_n$. Moreover, the roots of the 
  Legendre polynomial are symmetric with respect to zero. The $\C_k$ are thus symmetric with respect 
  to $1/2$ and  so will be the $c_k$ and $d_k$. The symplectic  integrator is  thus 
  symmetric, and this hypothesis was not necessary. This is not the case for Lie algebra
  symbolic computation, where the assumption that the integrator is symmetric decreases in a large 
  amount the  number of the variables.
For an integrator of type
\be
\SB_n :  \e^{d_1 V} \e^{c_2 U}\e^{d_2 V} \dots \e^{c_n U}\e^{d_n V}\e^{c_{n+1} U}\e^{d_{n+1} V} \ ,
\ee
we need to  set $\C_1=c_1 =0$ in   formula (\ref{eq.66}), which means that in the integration formula,
one node is fixed to an extremity of the interval $[0,1]$. 
On the other hand, we have $ \C_{n+1} =\sum_{i=1}^{n+1} c_i =1$. The problem is thus to find 
nodes and weights for a Gauss formula with fixed nodes at the boundary of the interval 
of integration. The solution is given by the Gauss-Lobatto formulas (Abramovitz and Stegund, 1965).
For $k=2,\dots,n$, we have $ \C_k =  (1+x_k)/2$ where $x_k$  are the $n-1$ roots of $P'_n(x)$, and
\be
d_1 = d_{n+1} = \Frac{1}{n(n+1)} ;\quad d_k = \Frac{1}{n(n+1)(P_n(x_k))^2} ;
\quad \hbox{for}  \quad k=2,\dots,n \ .
\ee

As previously, the integrators are  symmetrical. These relations  thus allow 
us to obtain in a straightforward manner symplectic integrators  for perturbed systems at 
any order without the  need to solve algebraic equations which are 
difficult to handle at large orders. Moreover,  it provides a  demonstration that 
this solutions exists at any order, with positive coefficients $c_k,d_k$. 

\section{Numerical examples}

In this section, we will test the efficiency of the family  of integrators $\SA_n$ and $\SB_n$ 
on a simple pendulum example and on a planetary problem.
For the simple pendulum
\be 
H = \Frac{p^2}{2} + \ve \cos q
\ee
we apply directly the previous computations with $A=p^2/2$ and $\ve B=\ve \cos q$. 
For each value of the stepsize $\tau$,
we have measured the maximum difference between the energy at the origine and 
the computed energy along the trajectories, over a time $T=25000$. This comparison is 
performed for $\ve = 0.1$ and $\ve = 0.001$ (Fig. 1). For $\SA_n$ or  $\SB_n$, the 
logarithm of differences are  plotted versus  $\log(\tau')$, where $\tau' = \tau/n$. 
In such a way, as $n$ is the number of evaluations of 
$\exp(c\tau L_A)$ and $\exp(d \tau  L_B)$ for the given integrator, 
the integrators are compared at constant cost. As expected,  for sufficiently 
small stepsize, the residuals 
behave as $\tau^2 \ve^2$ for $n \geq 2$, and as  $\tau^2 \ve $ for the leapfrog 
integrator ($n=1$).   It is also clear that for small stepsize, 
nothing is really gained by increasing the order of the integrator ($n$), beyond
$n=2$. 

For large stepsize, this is not true, as the term $\tau^{n+2} \ve$  or 
$\tau^{n+3} \ve$  (see next section) is still dominant, and we observe an increase 
of the slope  with the order of the integrator, until unstabilities appear, 
probably due to the divergence of the remainders (it should be reminded that
if a stepsize of $1$ is used for $\SA_1$, a stepsize of $n$ is used for 
$\SA_n$). In most cases, $n=3$ or $n=4$ seems to be the best choices.

\begin{figure}[h]
\begin{center}
 \includegraphics*[scale=0.65]{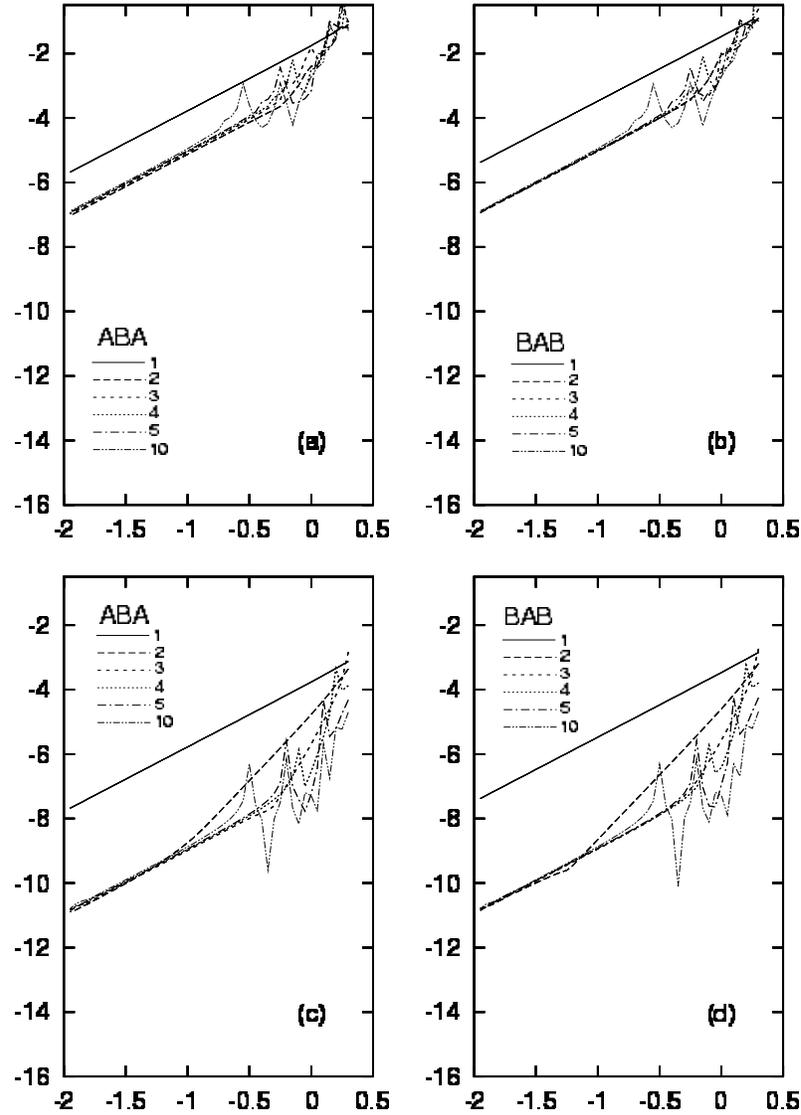}
  \caption{Fig.1--6. Logarithm of relative energy error plotted versus $\log(\tau')$, where $\tau'=\tau/n$,
 $\tau$ the stepsize, and $n$ is the index of the method (and  the curve)
 for the various symplectic integrators of the family $\SA_n$ and $\SB_n$.
 Fig. 1. Simple pendulum  with $\ve =0.1$ (a-b) and $\ve=0.001$ (c-d).}
  \label{fig1}
\end{center}
\end{figure} 

\begin{figure}[h]
\begin{center}
 \includegraphics*[scale=0.65]{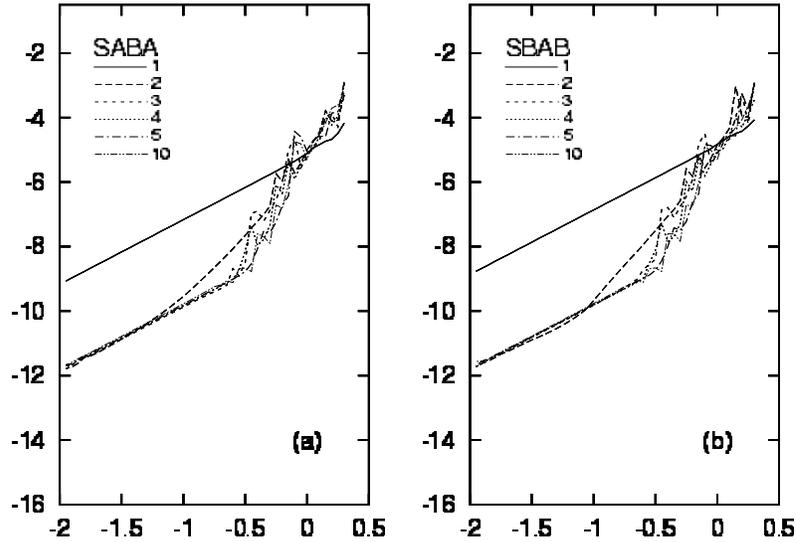}
  \caption{Relative energy error versus stepsize for the 
  Sun-Jupiter-Saturn problem in Jacobi coordinates for the family of integrators $\SA_n$ and $\SB_n$.}
  \label{fig2}
\end{center}
\end{figure}

In the case of the planetary $N$-Body problem, the situation is more complicated.
The Hamiltonian is splitted in an integrable Keplerian part, $A$, and a perturbation, $B$,
corresponding to the mutual gravitational  interaction of the planets.
The Keplerian part is integrated in elliptical coordinates, while the perturbation 
(which is essentially a sum of invert of the mutual distance of the 
planets) is integrated in rectangular cartesian coordinates (Wisdom and Holman, 1991).

There are several possible choice of coordinates for this decomposition. 
The initial choice of Wisdom and Holman, (1991), was to use Jacobi coordinates. 
In this case, $B$ is integrable, as it depends only on the positions  $q$. 
In Poincar\'e heliocentric coordinates (see Laskar and Robutel, 1995), the 
expressions are more simple, but the perturbation $B$ needs to be splitted 
in two terms $B=B_1(p)+B_2(q)$ which depends uniquely on the momentum $p$, 
or on the positions $q$. As the methods which are presented here depends only 
on the linear part (in $L_B$) of the integrator, 
they can be adapted in a straightforward manner to this case,  by 
substituting in their expresions $\exp L_{B_1} \exp L_{B_2}$ or  $\exp L_{B_2} \exp L_{B_1}$
to  $\exp L_{B}$. In doing so one needs to be sure that the final 
symplectic scheme is still symmetric, which will ensure that no terms of 
order 2 will appear in the decomposition of the corresponding formal Hamiltonian $K$
 in equation (\ref{eq.12}). The use  of these coordinates for symplectic integrators
 was first proposed by Koseleff (1993, 1996).
 
 In the present case, we will use Jacobi coordinates, as this choice will be motivated 
 by the next sections which require that $B$ depends only on $q$. In 
 Jacobi coordinates, we did the computation 
 for the Sun-Jupiter-Saturn system over 25000 years (Fig. 2), and obtained very similar results as 
 for the simple pendulum with $\ve =0.001$. This is understandable as this is of the order 
 of the ratio of perturbation due to the mutual interaction of the planets 
 over the potential of the Sun. 
 It can be clearly seen that for all $n\ge2$, these integrators outperformed by 
 several order of magnitude the precision of the leapfrog integrator, except for 
 very large stepsizes. The best choices being again $n=3$ or $n=4$.
 In all figures, it is very obvious that the $\tau^2\ve^2$ term  is the 
 main limiting factor. We will now make an explicit 
 computation of this term and present a strategy to get rid of it.
\section{Computation of the remainders}
We  compute here the remainders of the symplectic integrators 
$\SA_n$ and $\SB_n$.
By switching the role of $U$ and $V$ in (\ref{eq.gen_form}), we obtain easily  
\begin{prop}
Let $c_1,\dots c_n, d_1,\dots, d_n \in \bbbr$, such that $\sum_{i=1}^n c_i=\sum_{i=1}^n d_i = 1$. 
Then there exists $W \in \cL(U,V)$ such that
\be
\e^{c_1 U}\e^{d_1 V} \e^{c_2 U}\e^{d_2 V} \dots \e^{c_n U}\e^{d_n V}= \e^W
\label{eq.UV}
\ee
with
\be
\EQM{
W \tilde\equiv U  + V &+ \sum_{p=1}^{+\infty} \left(\sum_{k=1}^n d_k \,\Frac{B_p(\C_k)}{p!} \right) ad(U)^p\, V \crm  
&+ \sum_{p=1}^{+\infty}  \left(\sum_{k=1}^n c_k \,\Frac{B_p(\delta_{k-1})}{p!} \right) ad(V)^p\, U  
}
\ee
with $\delta_0=0, \delta_k=d_1+\dots+d_k$, and where $\tilde\equiv$ is the equivalence modulo terms of degree
$\geq 2$ in $U$ and $V$  in $\cL(U,V)$.
\end{prop}

If we apply this result to compute the largest  term in the remainder of the previous 
symplectic integrators,  we obtain for each integrator 
\be
\EQM{
W = A  + B &+ \left(\sum_{k=1}^n c_k \,\Frac{B_2(\delta_{k-1})}{2} \right)\{\{A,B\},B\}\tau^2\ve^2 \cr
&+\left(\sum_{k=1}^n d_k \,\Frac{B_p(\C_k)}{p!} \right) L_A^{2p}B\tau^{2p}\ve + O(\tau^4\ve^2+\tau^{2p+2}\ve) 
} 
\ee

We can be more specific for the two classes of integrators 
$\SA_n$ and $\SB_n$ by taking into account the fact that these integrators are 
reversible. In this case,  each integrator of the classes $\SA_{2n}$, $\SA_{2n+1}$, 
$\SB_{2n}$, $\SB_{2n+1}$,
with $\sum_{i=1}^n c_i=\sum_{i=1}^n d_i = 1$, is the time-$\tau$ evolution of the flow 
of the Hamiltonian $W$, with the following remainders :

\noindent
-- {$\SA_{2n}$}:
we have $2n+1$ steps with $d_{2n+1} = 0 $ and, for $p=0,\dots,n$
\begin{equation}
\left\{
\begin{array}{llll}
c_{n+1+p} &= c_{n+1-p} \ ;  \qquad   & \g_{n+p}  &= 1-\g_{n+1-p} \ ;\cr
d_{n+p} &= d_{n+1-p}   \ ;  \qquad    &\dd_{n+p}  &= 1- \dd_{n-p} \ ;\cr
\end{array}
\right.
\end{equation}
which gives, after reduction of the symmetries, and 
$d_n = 1/2 - \d_{n-1}$, $c_{n+1} = 1 -2\g_n
$
\be 
\EQM{
W = &A  + \ve B 
+ \left(\Frac{c_{n+1}}{2}B_2(1/2) + \sum_{k=1}^n c_k  B_2(\d_{k-1}) \right)\{\{A,B\},B\}\tau^2\ve^2 \cr
&+\left(2\sum_{k=1}^{n}  d_k B_{2n+2}(\g_k)  \right) \Frac{L_A^{2n+2}}{(2n+2)!}B\tau^{2n+2}\ve  
+ O(\tau^4\ve^2 + \tau^{2n+4}\ve) 
\label{eqw1}
}
\ee

\noindent
-- {$\SA_{2n+1}$}:
We have $2n+2$ steps, with $d_{2n+2}=0$, $d_{n+1} = 1-2\d_n$, $c_{n+1}= 1/2 -\g_n$, and, 
for $p=0,\dots,n$ 
\begin{equation}
\left\{
\begin{array}{llll}
c_{n+1+p} &= c_{n+2-p} \ ;  \qquad   & \g_{n+1+p}  &= 1-\g_{n+1-p} \ ;\cr
d_{n+1+p} &= d_{n+1-p}   \ ;  \qquad    &\dd_{n+p}  &= 1- \dd_{n+1-p} \ ;\cr
\end{array}
\right.
\end{equation}
which gives, after reduction of the symmetries 
\be
\EQM{
W &= A  +\ve B   
+ \left(\sum_{k=1}^{n+1} c_k  B_2(\d_{k-1}) \right)\{\{A,B\},B\}\tau^2\ve^2 \cr
&+\left(d_{n+1}\,B_{2n+4}(1/2)+2\sum_{k=1}^{n}  d_k \,B_{2n+4}(\g_k) \right) \Frac{L_A^{2n+4}}{(2n+4)!}B\tau^{2n+4}\ve \cr
&+ O(\tau^4\ve^2+\tau^{2n+6}\ve) 
\label{eqw2}
}
\ee

\noindent
-- {$\SB_{2n}$}:
This case is easily obtained by setting $c_1 = 0 $ in $\SA_{2n+1}$. We obtain for 
the new Hamiltonian

\be
\EQM{
W &= A  +\ve B  
+ \left(\sum_{k=2}^{n+1} c_k  B_2(\d_{k-1}) \right)\{\{A,B\},B\}\tau^2\ve^2 \cr
&+\left(d_{n+1}\,B_{2n+2}(1/2)+2\sum_{k=1}^{n}  d_k \,B_{2n+2}(\g_k) \right) \Frac{L_A^{2n+2}}{(2n+2)!}B\tau^{2n+2}\ve \cr
&+ O(\tau^4\ve^2+\tau^{2n+4}\ve) 
\label{eqw3}
}
\ee

\noindent
-- {$\SB_{2n+1}$}:
This case is easily obtained by setting $c_1 = 0 $ in $\SA_{2n+2}$.

\be
\EQM{
W &= A  + \ve B   
+ \left(\Frac{c_{n+2}}{2}B_2(1/2) + \sum_{k=2}^{n+1} c_k  B_2(\d_{k-1}) \right)\{\{A,B\},B\}\tau^2\ve^2 \cr
&+\left(2\sum_{k=1}^{n+1}  d_k B_{2n+4}(\g_k)  \right) \Frac{L_A^{2n+4}}{(2n+4)!}B\tau^{2n+4}\ve 
+ O(\tau^4\ve^2+\tau^{2n+6}\ve) 
\label{eqw4}
}
\ee

\section{Correctors}
\label{sec.cor}
The integrators $\SA_n$ and $\SB_n$  have very good properties for small values of the parameter 
$\ve$. Their numerical properties  were studied  in section 8. 
We have seen that the main limiting factor is  the term $\{\{A,B\},B\}$, which order is 
 $\tau^2\ve^2$. It would be of course very nice to get rid also of this term, 
 but the result of  Suzuki (1991) tells us
 that it is not possible to get rid simultaneously of the two terms $\{\{A,B\},B\}$ and $\{A,\{A,B\}\}$ with 
 integrators having  only positive values for the $c_i,d_i$ constants.
 It is not forbidden to have negative values for some of the constants, but 
 as $\sum c_i = \sum d_i =1$, having only positive constants ensures that the values of the 
 constants becomes smaller as the order of the integrator increases. This prevents  explosion of 
the  coefficients of the remainders
 which are polynomial in the $c_i,d_i$. 
 
 In order to get rid of the $\{\{A,B\},B\}$ term, one can use an alternate strategy, which 
is possible when $A$ is quadratic in the actions $p$, and $B$ depends only on the 
positions $q$ (this is in particular the case for the pendulum 
Hamiltonian, or for the $N$-Body problem when expressed in Jacobi coordinates). 
In this case, $\{\{A,B\},B\}$  depends only on $q$ and is thus integrable. It is then possible to compute 
it, and to add an additional step to the integrator $\cS$ of the form
\be
\cS_C =\e^{-\tau/2 c L_{\{\{A,B\},B\}}} \cS \e^{-\tau/2 c L_{\{\{A,B\},B\}}} 
\ee
where $c$ is the coefficient of $ \{\{A,B\},B\}$ in $W$ (Eq. \ref{eqw1}--\ref{eqw4}). The new corrected integrator $\cS_C$ is still
symmetric, and thus additional terms will appear only at order $\tau^4$. The values of the coefficients $c$ 
used in the correctors up to order 10 are listed in Table \ref{table_b}.  For some of the lowest orders, 
algebraic formulas can be given, but they become very rapidly cumbersome, and a better 
accuracy will be obtained by using the decimal value  which is given here with 40 digits.

\begin{table}[h]
\caption{Coefficients for the correctors up to order 10 for $\SA_n$ and $\SB_n$}
{\small
\begin{tabular}{ccc}
\trait
$n$ &$c_{\SA_n}$ & $c_{\SB_n}$ \\
\trait
 1 & $ 1/12   $                                      & $-1/24 $\\
 2 & $(2-\sqrt{3})/24  $ & $1/72 $ \\
 3 & $(54-13\sqrt{5})/648  $ & $ (13-5\sqrt{5})/288 $ \\
 4 & $0.003396775048208601331532157783492144$   &$(3861-791\sqrt{21})/64800$\\
 5 & $0.002270543121419264819434955050039130$   &$0.002381486672953634187470386232181453$\\
 6 & $0.001624459841624282521452258512463608$   &$0.001681346512091906326563693215296434$\\
 7 & $0.001219643912760418472579211822331645$   &$0.001251765616039400003072516100251191$\\
 8 & $0.000949308177745602234792177503535054$   &$0.000968797968073688571654684208462982$\\ 
 9 & $0.000759846022860436646358196674176815$   &$0.000772349023999952078227686810260323$\\
10 & $0.000621934331486166426497049845358646$   &$0.000630320044163167840798638762665112$\\
\trait
\end{tabular}
} 
\label{table_b}
\end{table} 

The plots of the residuals for these new integrators are presented  
in the case of  the pendulum  with $\ve=0.1$ and  $\ve=0.001$ (Fig.3-4), 
and the Sun-Jupiter-Saturn problem in 
Jacobi coordinates (Fig. 5). As we attain now the limitation due to round-off 
errors, computations were performed also in  quadruple precision. 
It is clear that now the slope of the residuals corresponds to the $\tau^4$ terms 
and that we got rid  of the $\tau^2\ve^2$ term.

\begin{figure}[h]
\begin{center}
 \includegraphics*[scale=0.65]{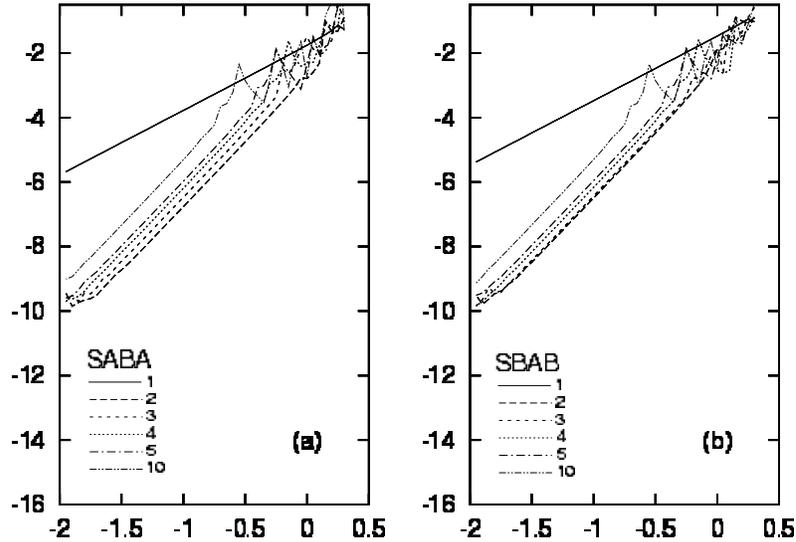}
  \caption{Relative energy error versus stepsize for the 
  simple pendulum with $\ve =0.1$ for   $\SA_n$ and $\SB_n$ with correctors.}
  \label{fig3}
\end{center}
\end{figure}

\begin{figure}[h]
\begin{center}
 \includegraphics*[scale=0.65]{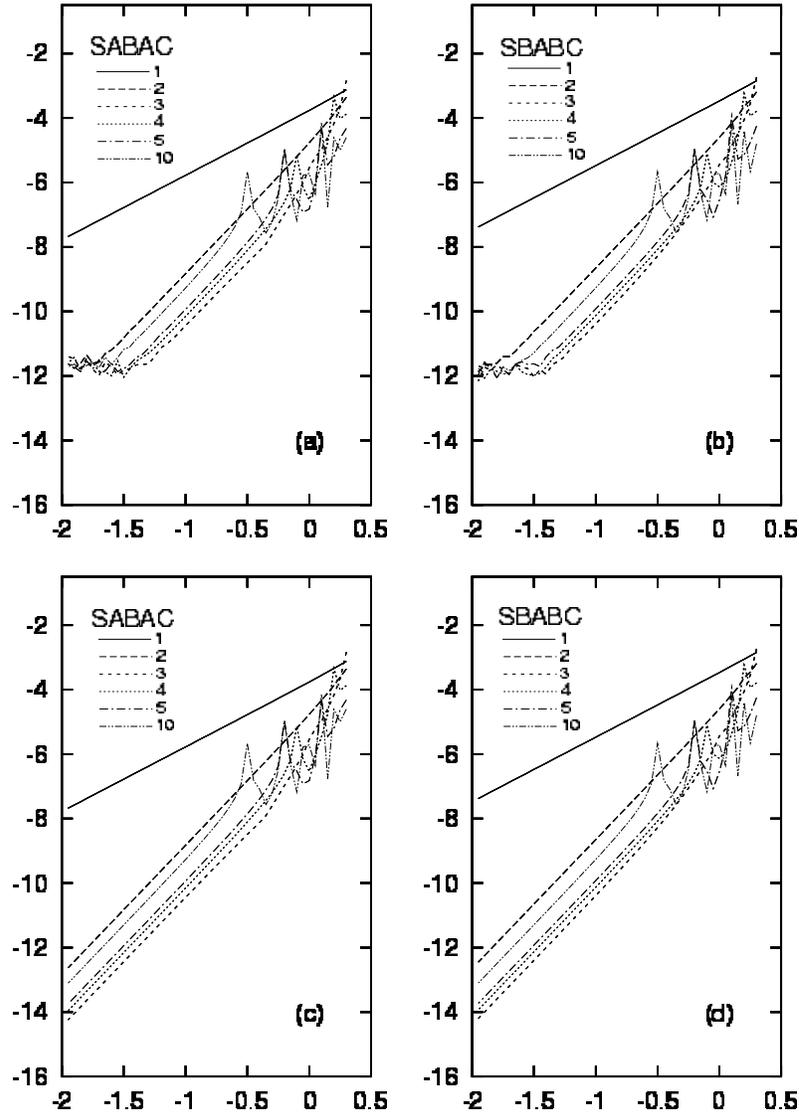}
  \caption{Relative energy error versus stepsize for the 
  simple pendulum with $\ve =0.001$ for   $\SA_n$ and $\SB_n$ with correctors in double (a-b)
  and quadruple (c-d)  precision.}
  \label{fig4}
\end{center}
\end{figure}

\begin{figure}[h]
\begin{center}
 \includegraphics*[scale=0.65]{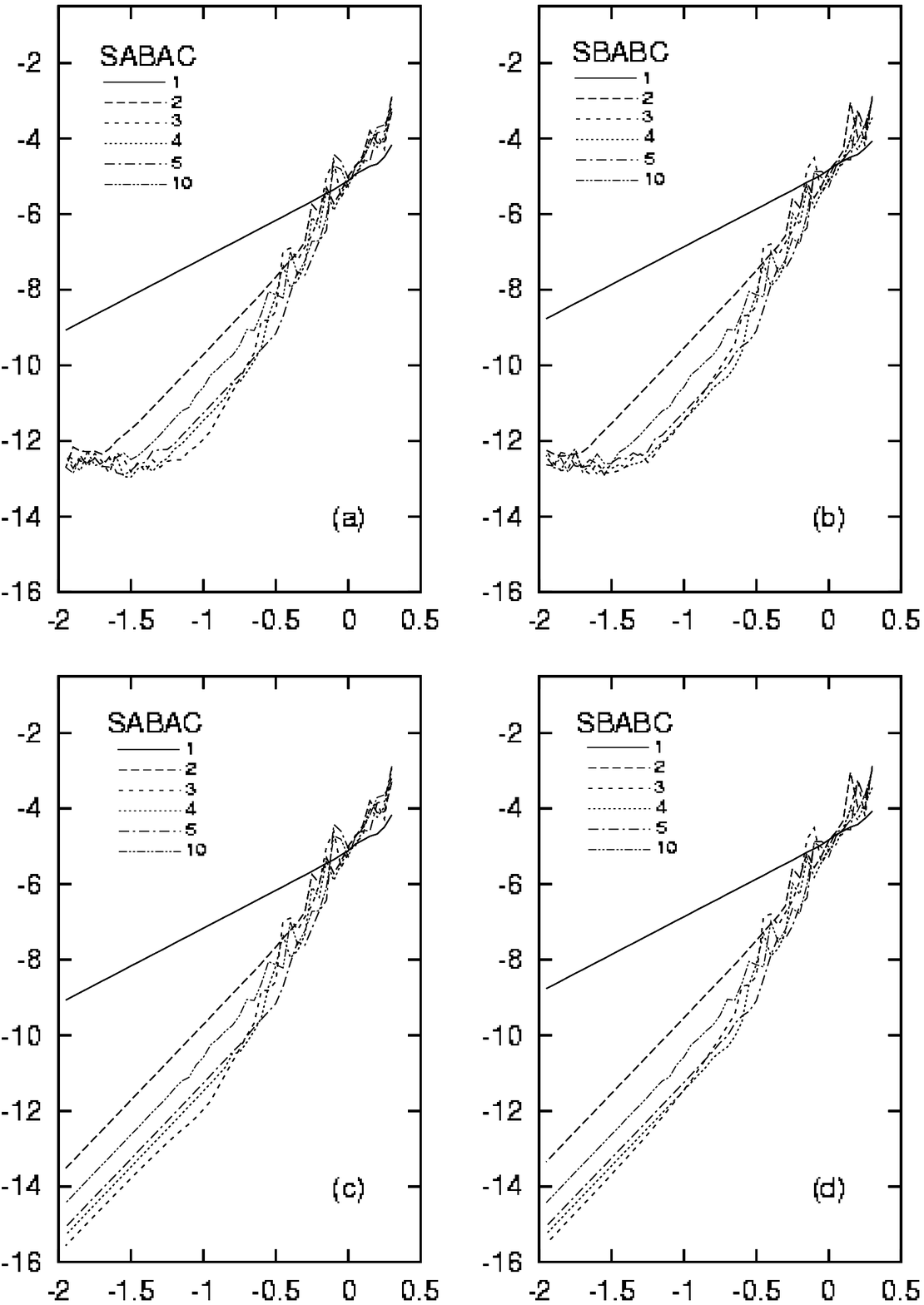}
  \caption{Relative energy error versus stepsize for the 
Sun-Jupiter-Saturn problem in Jacobi coordinates 
 for   $\SA_n$ and $\SB_n$ with correctors in double (a-b)
  and quadruple (c-d)  precision.} 
  \label{fig5}
\end{center}
\end{figure}

\section{Composition of integrators}
The corrector method of section \ref{sec.cor} provide a family of integrators 
$\SA_{Cn}$, $\SB_{Cn}$ of order 4 in $\tau$ and higher order in $\ve$  with remainders 
$O(\tau^4\ve^2)+O(\tau^k \ve)$ with $k = n+2$ for $n$ even, and $k=n+3$ for $n$ odd.
These integrators have very good  numerical properties, but it is still possible to 
improve them by using the composition method of Yoshida (1990). 
Indeed, if $\cS(\tau)$ is an integrator of order $2k$, then it is possible  to find 
$c$ such that 
\be
\cS(\tau) \cS(c\tau) \cS(\tau) 
\ee
is an integrator of order $2k+2$. Indeed, the symmetry of the integrator 
ensures that there are no terms in $\tau^{2k+1}$ in the remainders, 
and a straightforward computation 
gives  the condition of cancellation of the terms in $\tau^{2k}$
\be
c^{2k+1}+2 = 0
\label{cond.c}\ee 
that is
$ 
c = - 2^{\frac{1}{2k+1}}
$.
It should be noted that as $c$ is close to $-1$, the cost of this composition scheme, which we will denote $\cS^2$,
is roughly 3 times more expensive than the initial integrator $\cS$. Practically, we do one step forward, 
one step backward, and then one step forward again.
Nevertheless, if one generalises this sheme to a composition $\cS^{2m}$  defined as
\be
\cS^{2m}(\tau) = \cS^m(\tau) \cS(c\tau) \cS^m(\tau) 
\label{eqm} 
\ee
the condition (\ref{cond.c}) gives 
$
c = - (2m)^{\frac{1}{2k+1}} 
$. Usually $c$  is still not  very large, and the additionnal backward step  becomes negligeable 
for  large  values of $m$. 
Unfortunately, as one would expect, when $m$ increases, the size of the remainders also increases
and  when we analyse the  precision versus  cost,  it appears that we gain only  for small values of $m$ (Fig. 6).
These integrators are still interesting, especially when one searches for 
high accuracy, which means small step size. 

\begin{figure}[ht]
\begin{center}
 \includegraphics*[scale=0.65]{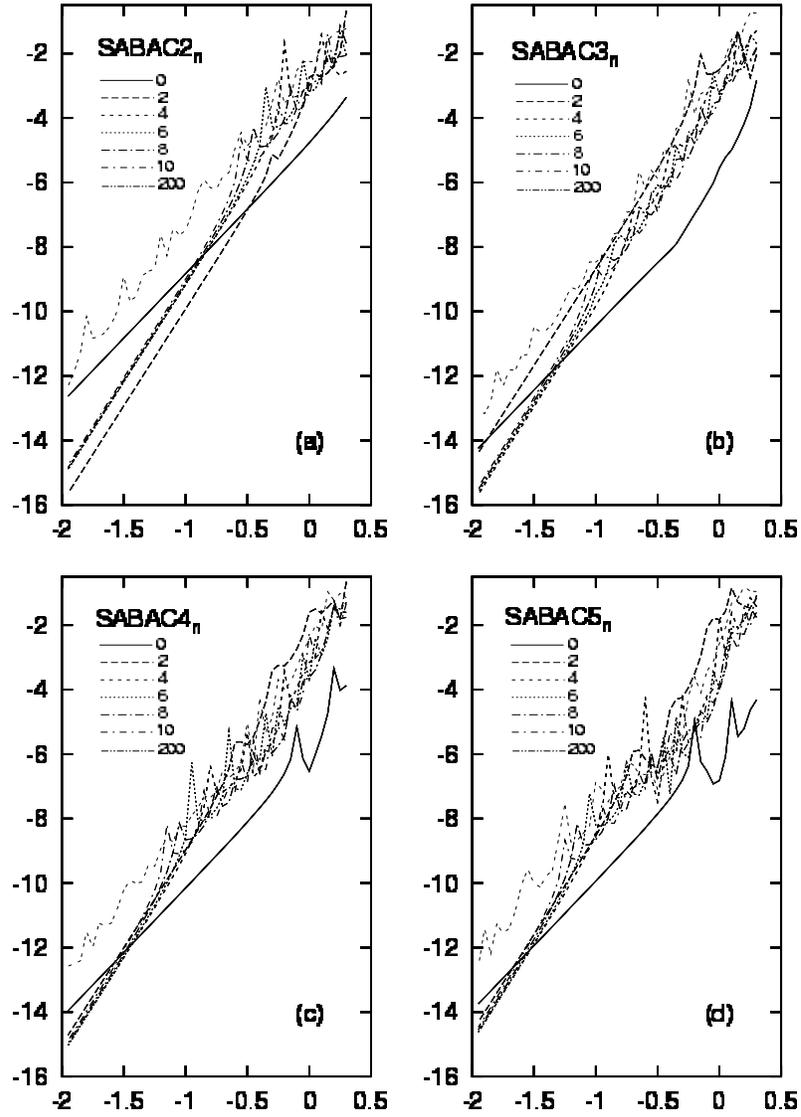} 
  \caption{Relative energy error versus stepsize for the 
  simple pendulum with $\ve =0.001$ for the  composition of $\SA_n$ and $\SB_n$ for 
  $n=2$ (a), $n=3$ (b), $n=4$ (c), and $n=5$ (d). The index of the curve corresponds 
  to the number of iterates $2m$  in the composition method (Eq.\ref{eqm}).}
  \label{fig6} 
\end{center}
\end{figure}

\section{Miscellanous remarks}

\subsection{Integrals}

The following result is obtained immediately:
\begin{prop}
Let $H = A+B$. If $F$ is an integral of $H$ and $F$ commutes with $A$ ($\{A,F\} =0$),
then $F$ is a true integral of  the symplectic integration of $H$ by any 
any of the integrators constructed above.
\end{prop}

Indeed, as $\{A,F\} =0$, and $\{H,F\} =0$, we have $\{B,F\} =0$, and thus 
$F$ commutes with any element of the free Lie algebra $\cL(A,B)$. Thus, if 
the integrator $\cS(\tau) $ is defined by 
$
\cS(\tau) = \e^{\tau L_W}
$
where $W \in \cL(A,B)$, we have $\{W,F\} =0$.
In particular, in Jacobi  or heliocentric  coordinates, the angular momentum 
depends only on the action variables and thus commute with the Hamiltonian of the 
Keplerian  unperturbed problem. The angular momentum is thus an exact integral 
of the symplectic integration of the $N$-body problem. In constrast, 
the initial Hamiltonian is only an approximate integral (at order $O(\tau^p\ve^2)+O(\tau^k \ve)$).
This feature can be used to check for the accumulation of errors in 
the integration.

\subsection{Non Hamiltonian systems}

In fact, the present results apply to general first order 
differential equations, and not only for Hamiltonian systems.  Indeed, 
the only properties which are used are formal properties of 
the Lie algebra of the Lie derivatives along the vector fields defined 
by $A$ and $B$.
If a differential system of order 1 can be written on the form
\be
\dot X = (L_A + L_B) X
\label{eq.sys}
\ee
where $L_A$ and $L_B$ are differential operators, for which
the two systems $\dot X =  L_A  X$ and $\dot X =   L_B  X$ are integrable, then the symplectic 
integrators defined above will apply in the same way. Even more, if $F$ is an integral of the 
system (\ref{eq.sys}) such that  $L_A\, F =0 $ and $L_B\, F=0$, then $F$ is also an integral for the 
symplectic integrator.

\subsection{$H=A+B_1 +B_2$}
It happens very often that the perturbation is not integrable, but can be splitted  in two parts 
$B=B_1+B_2$ 
which are integrable separately (this is the case in Poincar\'e heliocentric coordinates). 
As was already stated, the integrators $\SA_n$ and $\SB_n$ can be used provided some 
small modifications, but it will not be possible to use the correctors as defined in section 10.

\section{Conclusions}

We have presented here a new and constructive proof for the  existence at all orders
of the families of symplectic integrators $\SA_n$ and $\SB_n$, which were 
first described by McLachlan (1995).  We have also obtained the expressions  
of the leading terms of the remainders for all $n$. These integrators are 
particularly adapted to perturbed Hamiltonian systems of the form 
$ H = A + \ve B$, where $A$ and $B$ are integrable separately, and in particular for 
 planetary $N$-body problems.

Moreover, when $A$ is quadratic in the actions $p$ and $B$ depends only on the positions
$q$, the new family of integrators $\SA_{Cn}$ and $\SB_{Cn}$ given in section 10  provide integration scheme 
which is of order 4 in $\tau$, and has a remainder of the order of $O(\tau^4\ve^2 + \tau^p\ve)$,
where $p=n+2$ or $p=n+3$.
For practical use, it seems that the integrators for $n=3$ or $n=4$ are the most efficients.
Despite they require additional computations for the corrector, the corrected 
integrators  $\SA_{Cn}$ and $\SB_{Cn}$  will beat the simple integrators 
$\SA_{n}$ and $\SB_{n}$  in many occasions, but unless one search for very high accuracy 
with small stepsize, composition as described in section 11 is usually not  very useful.

All the integrators which are presented here have only positive stepsize, except for the
corrector. It should still be investigated whether some integrators of order 4 with 
negative stepsize could be useful.


\begin{acknowledgements}
We thank A. Albouy, A. Chenciner, D. Sauzin for very useful discussions,
and F. Joutel for his help in the implementation of the integrators.
Although it  does not appear in the final work, the development of these 
integrators was largely facilitated by the use of LIE\_TRIP, an algebraic 
manipulator for Lie algebra, which was developped with the unvaluable help of
M. Gastineau.
\end{acknowledgements}

\end{article}
\end{document}